\documentclass[aps,prc,nofootinbib,onecolumn,groupedaddress,showpacs,showkeys]{revtex4}

\usepackage{graphicx}
\usepackage{epsfig,latexsym,amssymb}
\usepackage{amsmath}
\usepackage{bm}
\usepackage{slashed}
\usepackage{subfigure}
\usepackage{comment}
\usepackage[dvipsnames]{xcolor}

\newcommand{\be}{\begin{equation}}
\newcommand{\ee}{\end{equation}}
\newcommand{\bea}{\begin{eqnarray}}
\newcommand{\eea}{\end{eqnarray}}

\newcommand{\as}{\alpha_s}

\def\eq#1{{Eq.~(\ref{#1})}}
\def\fig#1{{Fig.~\ref{#1}}}
\newcommand{\ben}{\begin{eqnarray*}}
\newcommand{\een}{\end{eqnarray*}}

\begin{document}

\title{Bulk quantities in nuclear collisions from
running coupling $k_{T}$-factorization and hybrid simulations}
\author{Andre V. Giannini, Fr\'ed\'erique Grassi and Matthew Luzum}
\affiliation{
Instituto de F\'{i}sica, Universidade de S\~ao Paulo, Rua do Mat\~ao
1371,  05508-090 S\~ao Paulo-SP, Brazil \\
}

\begin{abstract}
Starting from a Color Glass Condensate (CGC) framework, based on a
running-coupling improved $k_T$-factorized formula, we calculate bulk
observables in several heavy-ion collision systems.  This is done in
two ways:
first we calculate the particle distribution directly implied
from the CGC model, and we compare this to the case where it is instead
used as initial conditions for a hybrid hydrodynamic simulation.
%%%%
In this way, we can assess the effects of hydrodynamic and hadronic evolution
by quantifying
%,
how much they change the results from a pure initial state approach
and, therefore,  to what extent initial condition models can be directly
compared to experimental data.
We find that entropy production in subsequent hydrodynamic evolution can increase multiplicity by as much as 50\%.
However, disregarding a single overall normalization factor, the centrality, energy, and system size dependence of charged hadron multiplicity
is only affected at the $\sim$5\% level.
Because of this, the parameter-free prediction for these dependencies gives reasonable agreement with experimental data whether or not hydrodynamic evolution is included.  On the other hand, our model results are not compatible with the hypothesis that hydrodynamic evolution is present in large systems, but not small systems like p-Pb, in which case the dependence of multiplicity on system size would be stronger than seen experimentally.
Moreover, we find that hydrodynamic evolution significantly changes the distribution of momentum, so that
observables such as mean transverse momentum are very different from the initial  particle production,
and much closer to measured data.
%%%%
Finally, we find that a good agreement to anisotropic flow data cannot be achieved due to the large eccentricity generated by this model.
\end{abstract}

\maketitle
\vspace{1cm}

\section{Introduction}

Current colliders operating at ultra-relativistic
energies --- the Relativistic Heavy-Ion Collider (RHIC)
and the Large Hadron Collider (LHC) --- are designed
to study the behaviour of nuclear matter under extreme
conditions.
%%%
%%%
%%%
%%%
The matter formed right after a
high energy collision is thought to be a system out of equilibrium with
a large gluon occupation number, so each gluon carries a small fraction of
momentum, $x \ll 1$, of the original hadron~\cite{CGC.review}.
This implies that the knowledge of the properties of the small-$x$ modes
of the hadronic wave function is very important to understand the initial
stages of hadronic collisions at high energies.
Over the years much effort has been devoted to do so, and
it is now well established theoretically that such properties can be
described in terms of the Color Glass Condensate (CGC) effective
field theory~\cite{CGC.review,CGC.Raju.McLerran,CGC.effective.theory}.
Among other important features, the CGC encodes non-linear
dynamics and effects of the parton
saturation\footnote{The saturation of the partonic density
inside a hadron is a direct consequence of the well known steep
increase of the gluon density with lowering
$x$~\cite{CGC.effective.theory,parton.saturation} that is driven
by the gluon emission process, $g\rightarrow gg$. In simple terms,
it can be understood as the inclusion of the gluon recombination
process ($gg\rightarrow g$) which starts to be non-negligible due
to the high density of gluons in the hadronic wave function.}
phenomena,
which restore the unitarity of the scattering matrix
and is characterized by a dynamical scale $Q_{s}$, the saturation
scale, considered to be the typical momentum scale in the hadronic
wave function. The presence of this scale, which increases with
the energy of the collision and the atomic number, allows to
treat particle production on a solid basis where perturbative
methods can be applied.
%%%

After the initial particle production, the system can continue to interact and evolve.
In collisions between heavy nuclei (and possibly in smaller
systems~\cite{Khachatryan:2010gv,CMS:2012qk,Abelev:2012ola,Aad:2012gla,Adare:2013piz,
Adare:2014keg,Aidala:2017ajz,PHENIX:2018lia,Dusling:2015gta,Weller:2017tsr,Nagle:2018nvi}),
after a short period of time $\tau < 1$ fm/$c$, the system is believed to behave as a relativistic fluid.
Indeed, viscous hydrodynamic simulations have been quite successful at describing and predicting
various experimental data~\cite{Shen:2011eg,Bozek:2011if,Qiu:2011hf,Qiu:2011iv,Schenke:2012wb,Bozek:2012qs,Gale:2012rq,Qiu:2012uy,Gale:2013da,Schenke:2014zha,Noronha-Hostler:2015uye,Ryu:2015vwa,Niemi:2015qia,Shen:2015qta,deSouza:2015ena,Giacalone:2017dud}.

Particle number is intimately related to entropy, and in the limit of
ideal hydrodynamics (i.e., zero viscosity), entropy is conserved
during the evolution of the system.  Dissipative effects (from viscous evolution
as well as any non-hydrodynamic process such as the later decay of resonances)
break this conservation.  Nevertheless, because of this expected approximate entropy
conservation, it is common to directly compare particle distributions in the initial
state with experimental data on bulk quantities like total multiplicity and its dependence
on rapidity, centrality, collision energy, and collision system.  This allows one to
quickly gauge the success of theoretical models of particle production, under
the assumption that subsequent evolution of the system will not change
these bulk quantities.

There exist comparisons between
CGC and hydrodynamics calculations, with both sharing
the same initial state
dynamics~\cite{Hirano:2005xf,Hirano:2007xd,Song:2010mg}.
However, the comparisons are limited, for example involving
only a single collision system and energy.
%,

In this work we compute global, bulk observables
obtained separately from a purely CGC model, and a hybrid (hydro + transport) model simulation that shares
the same initial state dynamics. Such procedure allows to
assess the effects of hydrodynamic and hadronic evolution,
and quantify to what extent initial condition models can be directly
compared to experimental data. Different from previous studies
we compare the results of both simulations for different energies,
from RHIC to LHC, and collision systems, including predictions
for O+O and Ar+Ar that may be part of the LHC
program in the future~\cite{Citron:2018lsq}.
%%%
In the next section we briefly present the ingredients of
each simulation and then the results following from each
one of them.

\section{CGC and hybrid simulations}

In the dilute-dense approximation,  a $k_{T}$-factorized expression
for inclusive small-$x$ gluon production in the scattering of two
valence quarks can be derived~\cite{Kovchegov:2001sc}.
%%%
This approximation is natural for asymmetric collisions such as $p$-A.
%On the other hand
Conversely, the applicability to symmetric A+A collisions at mid-rapidity is
not clear.
%questionable.
In the latter case, due to its increasing complexity, one expects factorization
breaking corrections which modifies the basic $k_{T}$-factorized expression.
Although such corrections have already been studied~\cite{Gelis:2008rw}
in the past, the magnitude of these corrections in the kinematical range
probed at the LHC is still unknown.
%%%

On the other hand, while the correct momentum distribution in A+A collisions
can still only be reliably obtained by means of ``dense-dense" calculations, which make
no use of such $k_{T}$-factorized formula~\cite{Blaizot:2010kh}, phenomenological
applications of the $k_{T}$-factorized expression~\cite{Kharzeev:2001gp,
Dumitru:2011wq,Levin:2011hr,Tribedy:2011aa,Dumitru:2018gjm} have been
able to correctly describe the centrality and energy dependence of the charged
hadron multiplicity.
%%%
This can be understood as an indication that for large nuclei and
high energies, these observables are mainly determined by the centrality
and energy dependence of the saturation scale, and might not be highly
affected by factorization breaking effects.
%%%
Following the success of previous works, we also apply a
$k_{T}$-factorized expression to obtain momentum integrated
quantities in A+A collisions.

Originally, such a $k_T$-factorized expression was derived in a fixed-coupling
approximation~\cite{Kovchegov:2001sc}. Corrections related to the running of
the QCD coupling were calculated in Ref.~\cite{Horowitz:2010yg}.
%%%
While the initial result was obtained for a fixed rapidity configuration,
the authors proposed the following generalization of how these
running coupling corrections would modify the leading order
expression in the presence of non-linear small-$x$ evolution\footnote{The
notation follows the one from ref~\cite{Horowitz:2010yg}:
${\bm k}$ denotes the transverse momentum of the produced gluon while
${\bm q}$ and ${\bm k}-{\bm q}$ are the ``intrinsic'' transverse momenta
from the gluon distributions.}:
\begin{eqnarray}\label{eq:rcktfact}
  \frac{d \sigma}{dy \, d^2 k \, d^2 b} \, = \, N\,  \frac{2 \, C_F}{\pi^2} \,
  \frac{1}{{\bm k}^2} \, \int d^2q \int d^2b'\,
       {\overline \phi}_{h_1} ({\bm q}, x_{1},{\bm b'})
  \, {\overline \phi}_{h_2} ({\bm k} - {\bm q}, x_{2},{\bm b'}-{\bm b})
  \,
  \frac{\as \left(
      \Lambda_\text{coll}^2 \, e^{-5/3} \right)}{\as \left( Q^2 \,
      e^{-5/3} \right) \, \as \left( Q^{* \, 2}\, e^{-5/3} \right)} \,\,,
\end{eqnarray}
where $N$ is an overall normalization to be fixed by
comparison with the experimental data, $C_{F}=(N_{c}^{2}-1)/2N_{c}$,
with $N_{c} = 3$, $x_{1,2} = (k_{T}/\sqrt{s}){\rm exp}({\pm y})$
is the momentum fraction of the projectile and the target
quark, respectively, and $\Lambda_\text{coll}^2$ is
a collinear infrared cutoff. Despite including higher-order corrections,
note that \eq{eq:rcktfact} is still $k_{T}$-factorized.
%%%%
The number of produced gluons with rapidity $y$ and momentum $k$ at
a coordinate ${\bm b}$ in the transverse grid in a given hadronic
collision $h_1 + h_2$ can be obtained from \eq{eq:rcktfact} as
%%%
\be\label{eq:Ng}
\frac{d N_g}{dy \, d^2 k \, d^2 b} = \frac{1}{\sigma_s} \frac{d \sigma}{dy \, d^2 k \, d^2 b} \,\,,
\ee
with $\sigma_{s}$ representing the effective interaction area of the hadrons $h_{1}$ and $h_{2}$.

In the above equation, ${\overline \phi}_{h_{i}} ({\bm k}, x,{\bm b})$
denotes the unintegrated gluon distribution (UGD), which
represents the probability of finding a gluon with momentum fraction
$x$ with transverse momentum $k_{T}$ in the hadron $h_{i}$~\cite{Albacete:2012xq}.
This distribution can be expressed as~\cite{Horowitz:2010yg}
\begin{equation}\label{eq:rc_ktglueA}
  {\overline \phi} ({\bm k}, y,{\bm b}) = \as\, \phi ({\bm k}, y,{\bm b}) = \frac{C_F}{(2 \pi)^3} \,
  \int d^2 r \, e^{- i {\bm k} \cdot {\bm r}} \ \nabla^2_r \,
  \mathcal{N}_{A} ({\bm r}, y,{\bm b})\,,
\end{equation}
%%%
with $\mathcal{N}_{A} ({\bm r}, y,{\bm b})$ denoting the
forward dipole scattering amplitude in the adjoint
representation at impact parameter $\bm b$. Although
some advances have been made very recently~\cite{Cepila:2018faq},
computing the matter distribution in $\bm b$ space
inside a proton directly from the CGC framework is
still an open and non-trivial problem~\cite{GolecBiernat:2003ym}.
Due to this limitation, a uniform gluon density within the proton
has been assumed; in this case, the integration over $d^{2}b'$ in
\eq{eq:rcktfact} generates a factor proportional to $\sigma_{s}$ which
cancels out with the same factor in the denominator of \eq{eq:Ng}.
%%%
Moreover, %%%%%%
$\mathcal{N}_{A}$ will be given by solutions of the
running coupling Balitsky-Kovchegov (rcBK)
equation~\cite{rcBK} provided by the AAMQS fit
of the HERA data on lepton-hadron collisions~\cite{Albacete:2009fh}.
Here we consider their solution with the
McLerran-Venugopalan model~\cite{Albacete:2009fh}
as initial condition. We note however that results for bulk
observables do not differ much when considering
other UGD sets which employ different
initial conditions for the rcBK evolution~\cite{Dumitru:2018yjs}.

Apart from incorporating running coupling corrections, the
other novel feature of \eq{eq:rcktfact} is the fact that all the
scales present in the $\as$ factors are fixed and determined by
explicit calculations\footnote{In contrast, in the fixed coupling
expression, the $\as$ factors in the denominator of \eq{eq:rcktfact}
are part of the UGD definition, given by the first equality in
\eq{eq:rc_ktglueA}, and have to be fixed by hand.}. We refer
to~\cite{Horowitz:2010yg} for the full expression
of the $Q^{2}$ dependence for the two $\as$ factors appearing
explicitly in \eq{eq:rcktfact}. A comparison of the centrality
dependence of charged hadron multiplicity in p+Pb and Pb+Pb
collisions~\cite{Dumitru:2018gjm} shown a difference of
$\sim$10\% in the results from \eq{eq:rcktfact} and the fixed
coupling $k_{T}$-factorization formula with the momentum scales
figuring in the $\as$ factors fixed by hand.

\eq{eq:rcktfact} is the starting point for each calculation.   In the case where hydrodynamic evolution is absent, one must still convert this spectrum of gluons into that of the hadrons which are measured.   This can be done by a convolution with a fragmentation function that represents the hadronization process.
%
%%%
By doing so, one also fixes %%%  the collinear infrared cutoff
$\Lambda_{\rm coll}^{2}$, as it should match the
momentum scale figuring in the fragmentation
function~\cite{Kovchegov:2007vf}, (which is usually chosen
to be proportional to the transverse momentum of
the produced hadron, $\mu_{FF}^{2}\sim p_{T}^{2}$).
%%%
While \eq{eq:rcktfact} implicitly assumes the validity of
collinear fragmentation functions to convert gluons into
hadrons~\cite{Horowitz:2010yg,Kovchegov:2007vf},
these ingredients have important limitations on
their range of applicability, being restricted to large momenta,
usually above 1 GeV.
%%%
Since bulk observables, as the ones we are interested here, have
significant contribution below this regime, they miss most
of the dynamics encoded in these fragmentation functions.
%%%
Because of this, we use the Local-Parton-Hadron-Duality
(LPHD)~\cite{Dokshitzer:1991eq}
as fragmentation model, where distributions at partonic and
hadronic level only change by a constant multiplicative
factor.

This same setup has already been considered
in~\cite{Duraes:2016yyg} and~\cite{Dumitru:2018gjm} where
\eq{eq:rcktfact} has been employed to obtain, respectively,
qualitative and quantitative results for bulk observables
in the CGC approach. Here, the pure CGC simulation follows
the one of~\cite{Dumitru:2018gjm}, which we extend to
the calculations of the average transverse momentum.
%%%
%%%
Moreover, we present predictions for
the centrality dependence of the charged hadron multiplicity that
may be measured in other collision systems (Ar+Ar and O+O) which
were not considered before.
%%%
The use of the LPHD is equivalent to disregarding any medium
or dynamical effects during the evolution of the system
created after the collision nor after the transition from a
state of deconfined matter to hadrons. This approximation will
be contrasted to the results from a more complete simulation of
heavy-ion collisions where such medium effects and the
dynamics at hadronic level are accounted for.

In the case   where collisions are described via a hybrid model,
the initial conditions for hydrodynamic evolution consist of the energy momentum tensor $T^{\mu\nu}$
at some initial time.    Since we are mainly interested in mid-rapidity observables, we perform
boost-invariant simulations, with initial conditions based on the distribution
of gluons at zero rapidity.  Specifically, we take the entropy density
to be proportional to the gluon density from the CGC framework
\be
s({\bm b},\tau = \tau_{0})
%= \frac{1}{\tau_{0}}\frac{dS}{d^{2}b\,deta}\bigg|_{y=0}
\propto
%\frac{1}{\tau_{0}}
\frac{dN_{g}}{d^{2}b\,dy}\bigg|_{y=0} \,,
%%%% \quad\quad  \frac{dN_{g}}{d^{2}b\,dy} \sim \int\,  d^{2}k\, [\, {\rm Eq.}\, 2\,] \,.
\ee
where the $dN_{g}/d^{2}b\,dy$ can be obtained by integrating
\eq{eq:Ng} over $d^{2}k$.
%%%
The corresponding energy density is then obtained by thermodynamic relations
from an equation of state derived from Lattice QCD calculations, s95p-v1.2 \cite{Huovinen:2009yb}.  We assume zero
initial shear tensor and bulk pressure, and no initial transverse fluid velocity.
%%%

In the above expression, $\tau_{0}$ represents the time at which the system starts behaving
hydrodynamically. As we do not include any pre-equilibrium description of the system and also
only account for the diagonal terms of the energy-momentum tensor
we assume that early/fast thermalization happens so the produced system
can start expanding in all directions as it should.
%%%
The results presented in the next section have been obtained using
$\tau_{0}= 0.2$ fm; the proportionality constant figuring in
$s_{0}({\bm b})$ will be fixed through the same experimental
data used to fix the overall normalization in the pure CGC
simulation.

Here we consider the cases where the system evolves hydrodynamically with and
without the presence of dissipative corrections in $T^{\mu\nu}$.
%%%
The resulting equations of motion in the dissipative case are
the ones from the second-order viscous hydrodynamics presented in~\cite{Paquet:2015lta}.
%%%
Those are solved using the MUSIC code~\cite{Schenke:2010nt}.
The cessation of hydrodynamic evolution is described by switching to the
hadronic afterburner UrQMD~\cite{UrQMD} once the system has locally reached
a chosen switching temperature $T_{sw}$.

We choose hydrodynamic parameters ($\eta/s(T), \zeta/s(T), T_{sw}$) to correspond to
the maximum a posteriori parameters from a comprehensive Bayesian
analysis~\cite{Bernhard:2018hnz}\footnote{This reference presents the result of several
analyses. The values considered
here are the ones quoted in table 5.8, corresponding to the most recent.}.
That analysis used different initial conditions,
and therefore these parameters are not necessarily the choices
that will give the best fit to experimental data for our initial conditions.
Nevertheless, they represent a reasonable
and realistic starting point.

The effects of dissipation are estimated by
completing a separate set of ideal hydrodynamic simulations, with exactly
the same set of initial conditions.

As seen above, the CGC is a natural choice of initial conditions.
Therefore we compare two different scenarios:
the pure CGC and the hybrid simulations (with CGC initial conditions).
As both simulations have a common starting point it allows for a more
consistent comparison.

In the next section we compare our results for the centrality and energy
dependence of charged hadrons produced in heavy-ion\footnote{The isotope
$^{129}$Xe has been used for collisions involving Xenon nuclei at LHC energies;
the parameters characterizing its deformation are the same from table II
of~\cite{Loizides:2014vua}.} and p+A collisions at
RHIC and LHC energies from the pure CGC (denoted as ``rcBK") and from the hybrid
model (with ideal or viscous hydrodynamics evolution; each case
is simply denoted by ``ihydro" and ``vhydro" respectively) simulations.

\section{Results and discussion}

\begin{figure}[htb]
\begin{center}
\includegraphics[width=8.5cm]{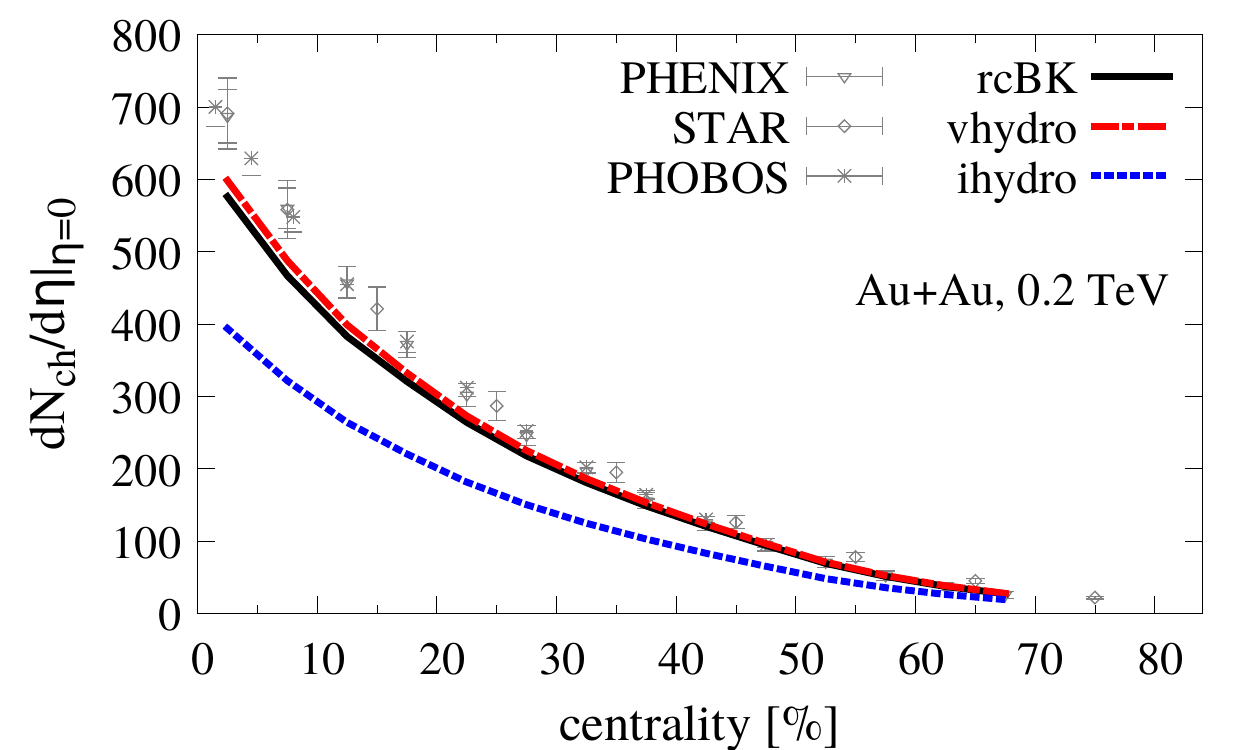}
\includegraphics[width=8.5cm]{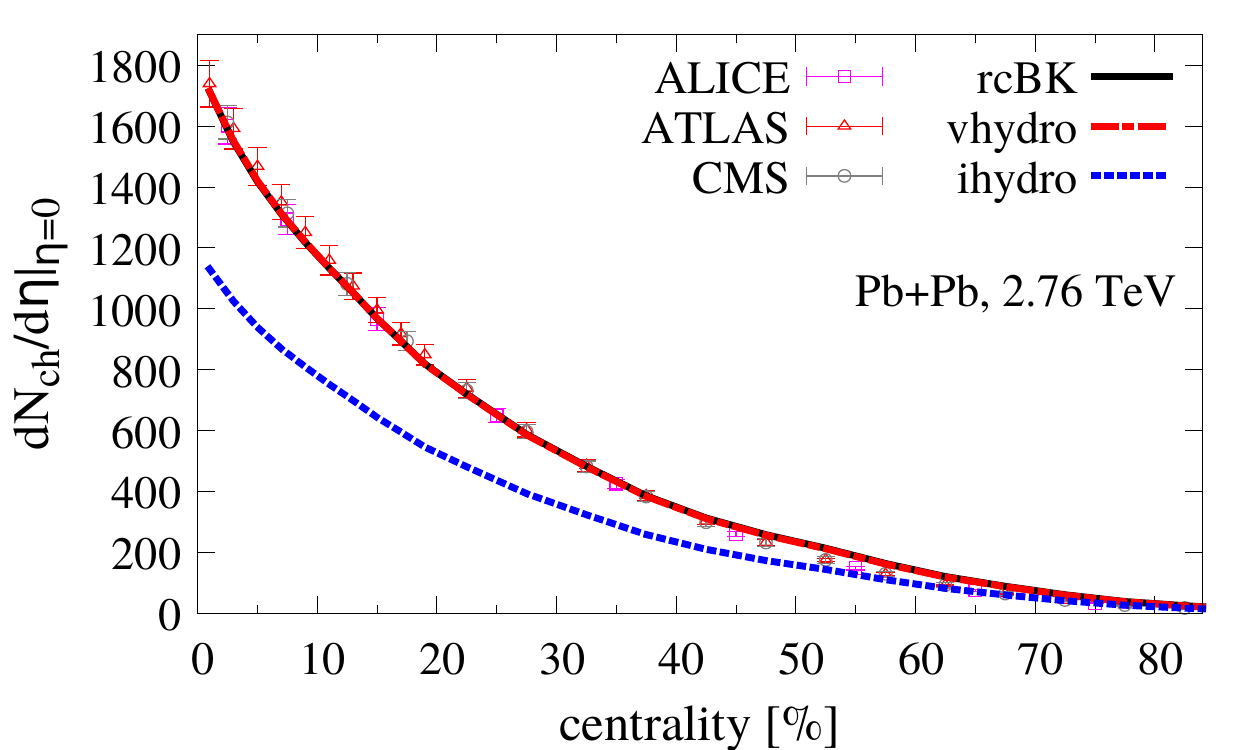}
\includegraphics[width=8.5cm]{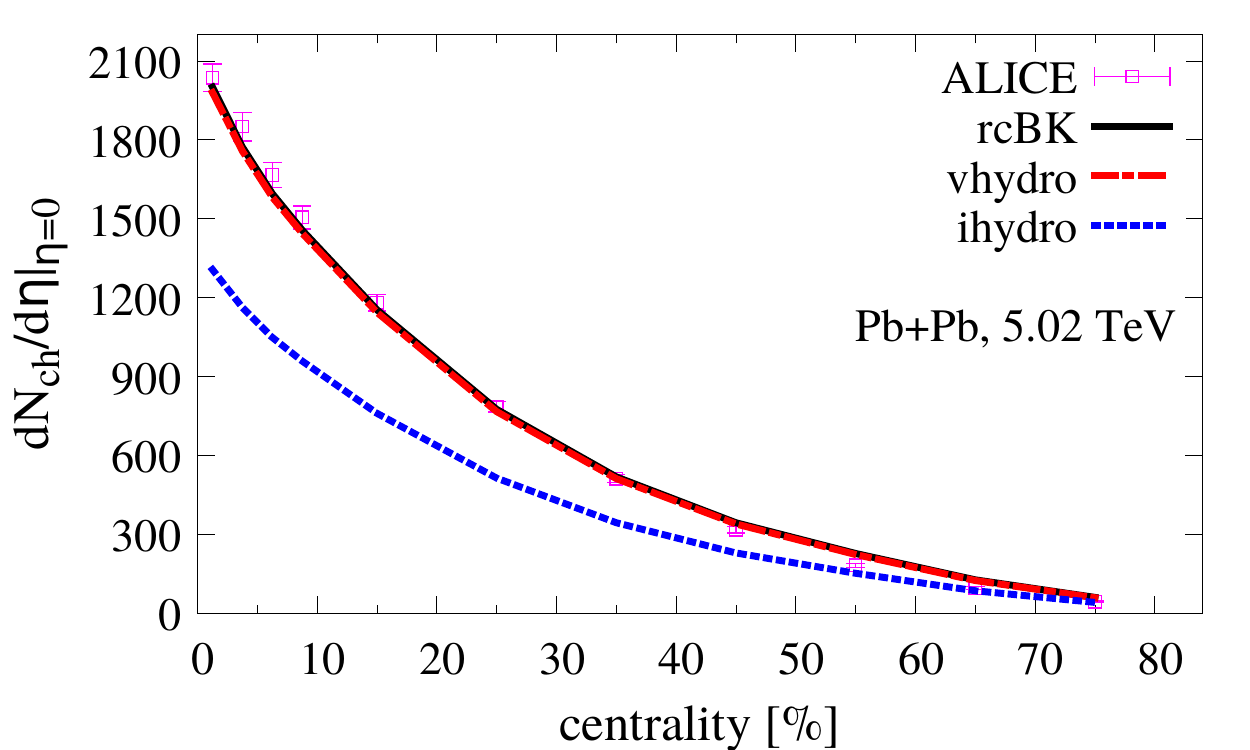}
\includegraphics[width=8.5cm]{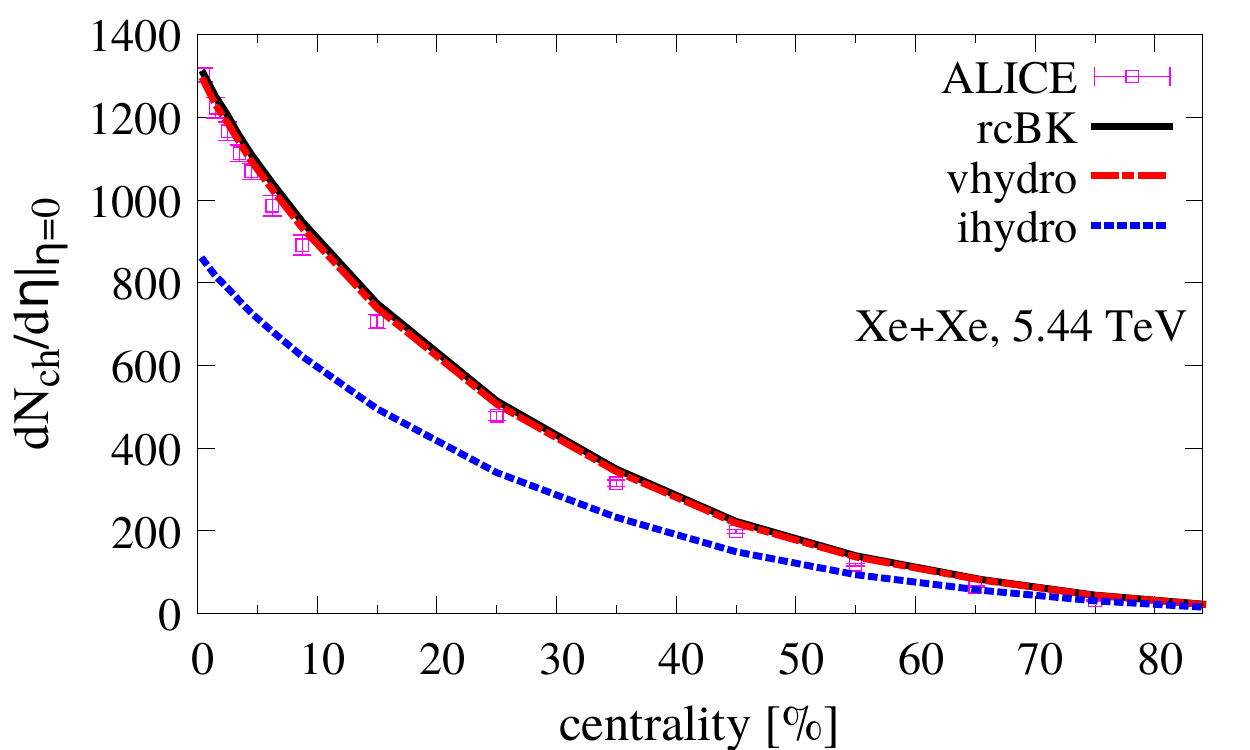}
\end{center}
\vspace*{-7mm}
\caption[a]{Centrality dependence of charged hadron multiplicity in the central
pseudorapidity region produced in A+A collisions at RHIC and LHC energies.
The experimental data is from~\cite{Adler:2004zn,Back:2002uc,
Abelev:2008ab,Chatrchyan:2011pb,ATLAS:2011ag,Aamodt:2010cz,Acharya:2018hhy}.
}
\label{fig:chpart_cent_dependence}
\end{figure}

\fig{fig:chpart_cent_dependence} shows the centrality dependence of the
charged hadron multiplicity produced in the central pseudorapidity region
of A+A collisions at RHIC top energy and also different LHC energies.
%%%
The normalization has been fixed by matching the rcBK
and the vhydro results to the data for central Pb+Pb collisions at
2.76 TeV. The normalization of the
viscous hydrodynamic
(vhydro) results has also been applied to
the ideal hydrodynamic (ihydro) ones. The same normalization is used across
all energies and collision systems.
%%%

We see that the pure CGC calculation gives a reasonable, but not perfect
fit to data. The hybrid calculation gives
similar results, showing that, indeed, the dynamics of the late
stages of nuclear collisions have only a small effect on how
multiplicity of charged hadrons is distributed across
centrality, energy, and collision systems.
Moreover, the poorest fit to data occurs at
the lowest collisions energies.

\begin{figure}[htb]
 \begin{center}
 \includegraphics[width=8.5 cm]{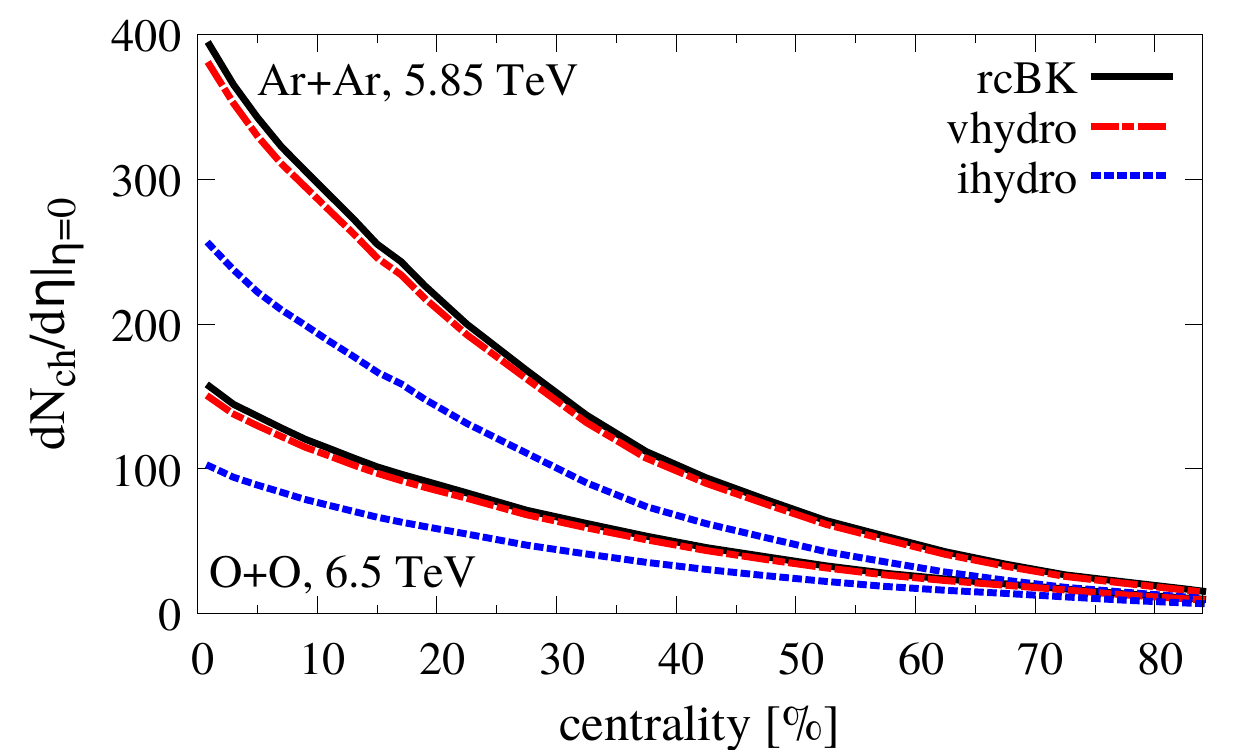}
\end{center}
\vspace*{-7mm}
\caption{Prediction for the centrality dependence of the charged hadron multiplicity
produced in Ar+Ar and O+O collisions.}
\label{fig:ArAr.OO.prediction}
\end{figure}

We also present predictions from the pure CGC and
hybrid simulations for the centrality dependence of the charged hadron
multiplicity produced in Ar+Ar and O+O collisions (as proposed for the LHC)
in \fig{fig:ArAr.OO.prediction}.

\begin{figure}[htb]
 \begin{center}
 \includegraphics[width=8.9 cm]{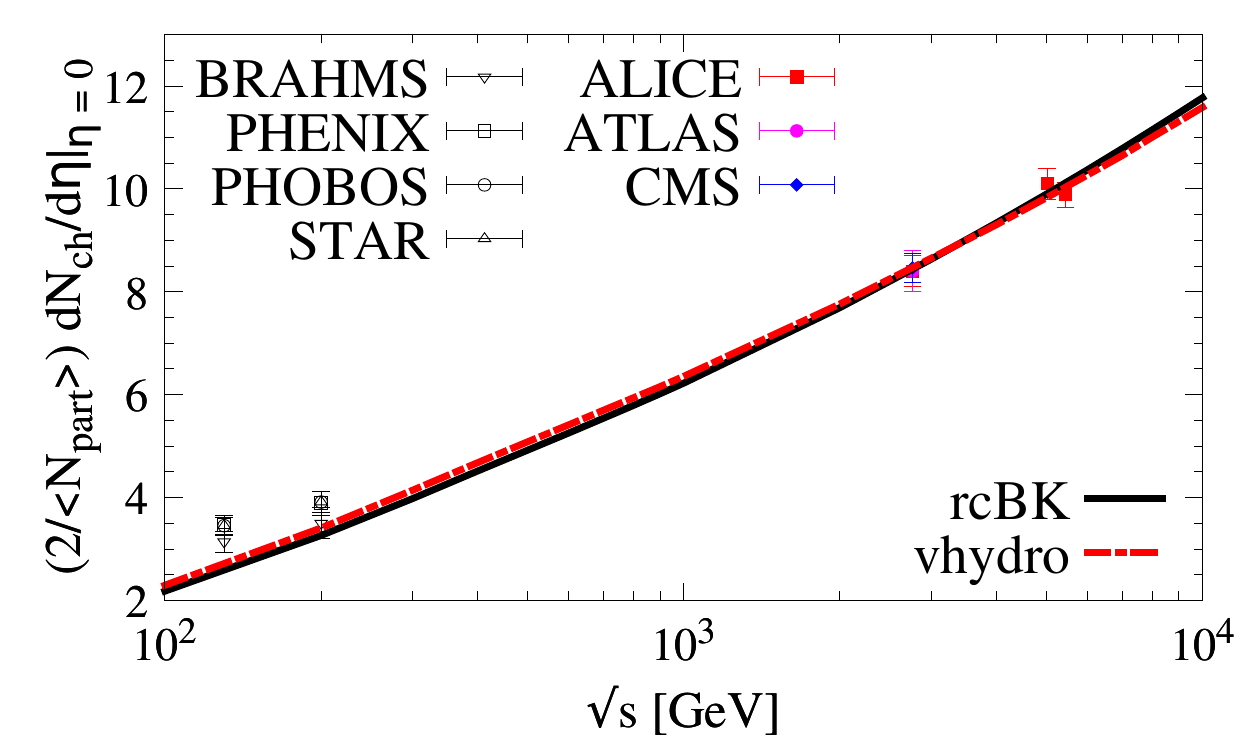}
 \includegraphics[width=8.9 cm]{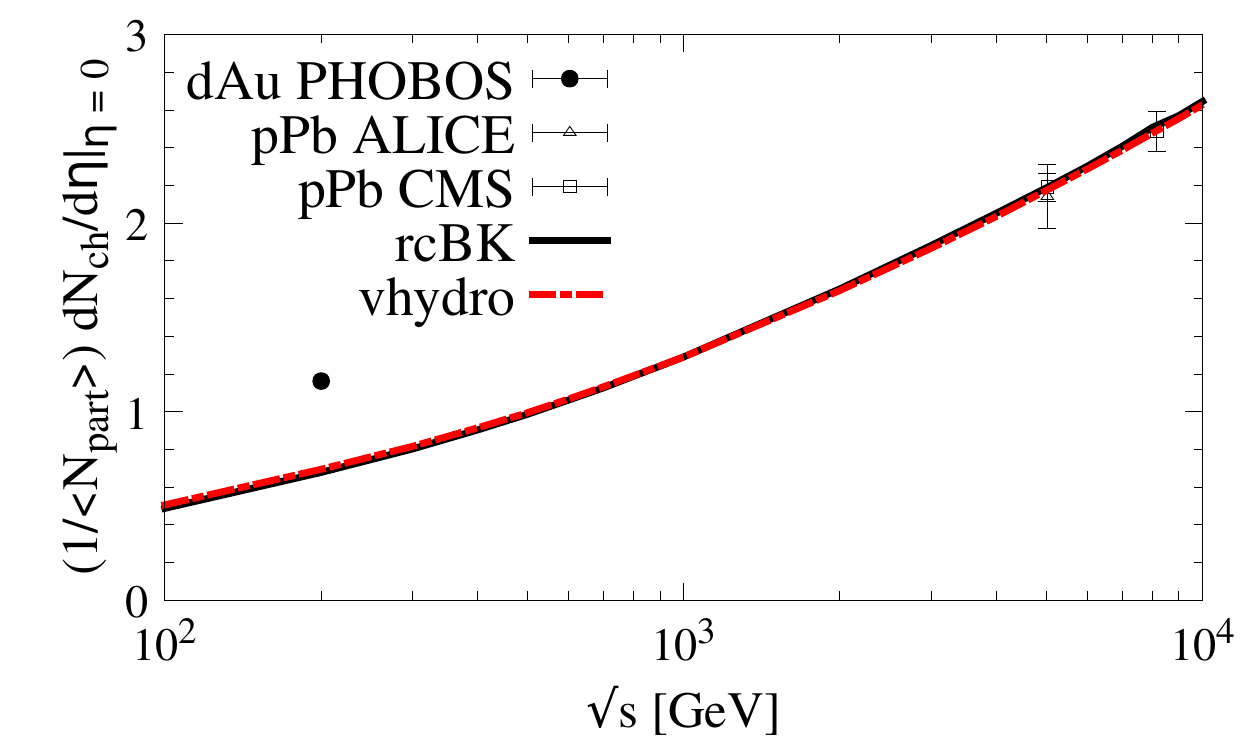}
\end{center}
\vspace*{-7mm}
\caption[a]{Energy dependence of the charged hadron multiplicity in (left) 6\% most central A+A collisions and (right) minimum bias p+Pb collisions. The experimental data is from~\cite{Adler:2004zn,Back:2002uc,
Abelev:2008ab,Chatrchyan:2011pb,ATLAS:2011ag,Aamodt:2010cz,Acharya:2018hhy,
Back:2003hx,ALICE:2012xs,Sirunyan:2017vpr}.}
\label{fig:sqrts_evolution}
\end{figure}

In \fig{fig:sqrts_evolution} we show the energy evolution
of the charged hadron multiplicity per participant (pair)
in p+A (A+A) collisions from the present hybrid simulation
and the rcBK results. %%% from~\cite{Dumitru:2018gjm}.
We can see that in A+A collisions, the pure CGC model predicts
a stronger energy dependence than is seen in experimental data.
Including hydrodynamic evolution counteracts this by
weakening the energy dependence, but the effect is
too small to achieve agreement with experimental data.

Similarly, the predicted energy dependence in light-heavy systems
also appears to be too strong.   Note that  the calculation presented in the right panel is for
p+Pb collisions, while the lowest energy experimental point is for a d-Au system.
However, we checked that there is no improvement
at $200$ GeV if we consider d+Au collisions instead.

Improving the rcBK results (and therefore the corresponding
initial condition for hydro simulations) at RHIC energies
for p(d)+A collisions requires a better knowledge of the
proton unintegrated gluon distribution at $x\sim 0.1$.
This will certainly have an impact on the results for A+A
collisions in the same energy range since nuclear distributions
are usually built from what is known about the proton's
structure function.
%%%
Such improvement will be possible in
the future as our knowledge about (un)integrated parton
distribution functions will be improved with the help of
an (so far planned) Electron-Ion collider.

\begin{figure}[htb]
\begin{center}
\includegraphics[width=8.5cm]{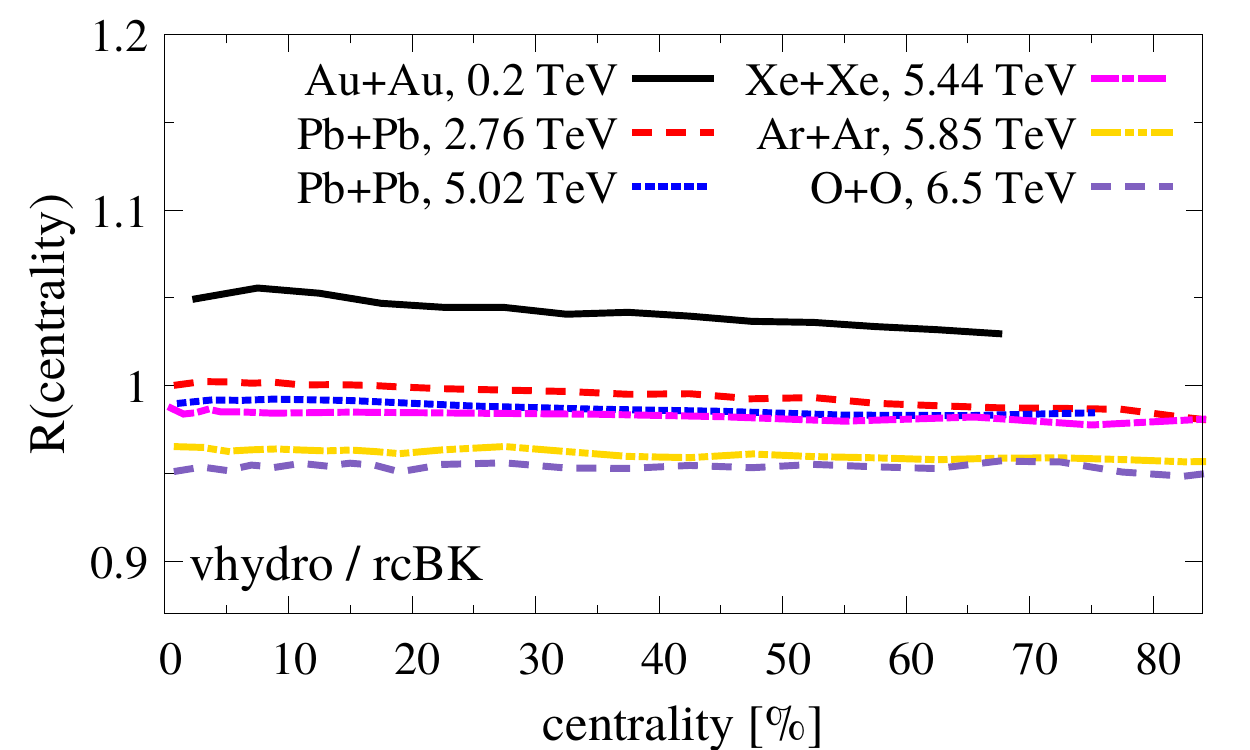}
\includegraphics[width=8.5cm]{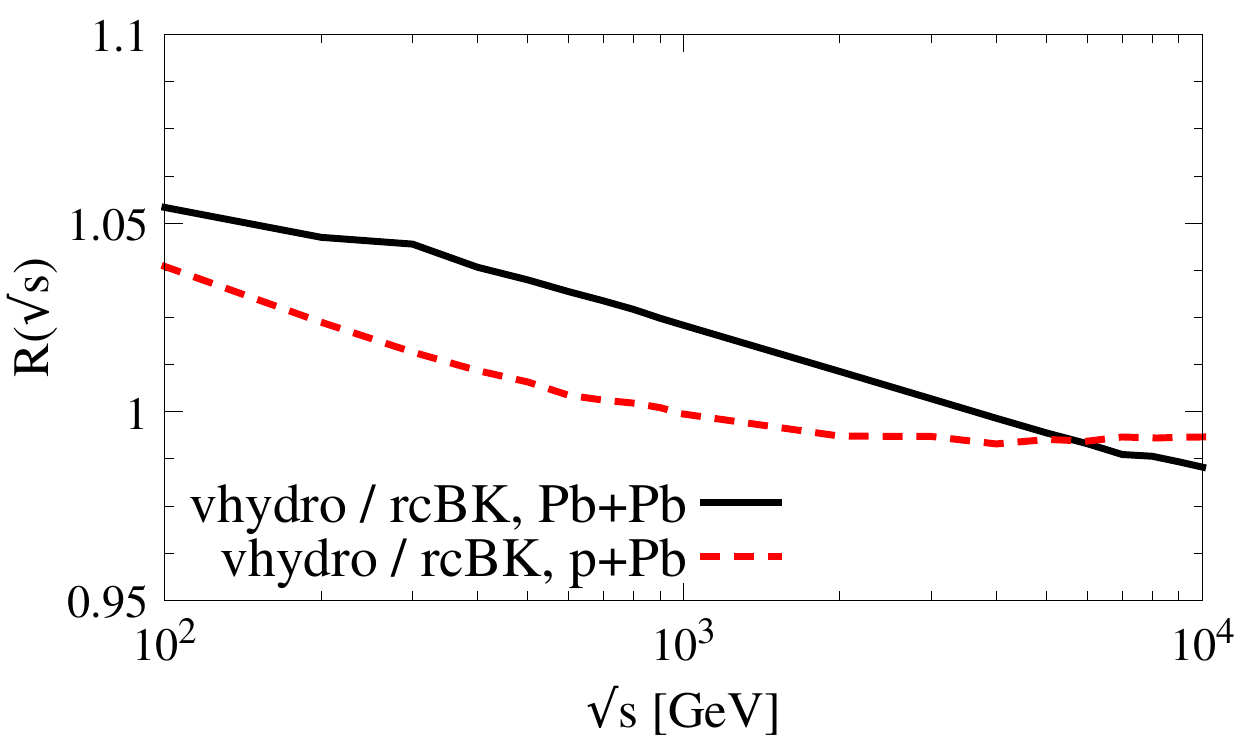}
\end{center}
\vspace*{-7mm}
\caption[a]{Centrality and energy dependence of the ratio of our results
with viscous hydrodynamics to the ones with only initial state dynamics for the charged
hadron multiplicities.}
\label{fig:ratio_viscous_rcBK}
\end{figure}

We can better quantify the effect of final-state evolution
by calculating ratios of charged hadron multiplicity in CGC
compared to hybrid calculations.
%%%
In \fig{fig:ratio_viscous_rcBK}, multiplicity ratios are presented as a
function of centrality in A+A and center of mass energy in
p+A and A+A collisions.

On the left, one can see that the ratio is almost constant as a function of centrality, showing no
more than a few percent change.   The largest differences come with
a change in collision energy and colliding system.  Recall that the normalization
factor was set from central Pb+Pb collisions at 2.76 TeV.  The biggest
difference in multiplicity comes in the system with the largest difference in
energy --- Au+Au at 0.2 TeV, where the vhydro multiplicity is $\sim$5\% higher
than rcBK.

The dependence on energy of the multiplicity ratio is shown explicitly on the right panel
of \fig{fig:ratio_viscous_rcBK} for the 6\% most central Pb-Pb collisions
as well as $p$-Pb collisions.  Across 2 orders of magnitude in collision energy,
hydro and final state effects change the predicted multiplicity by slightly more than 6\% for the A+A system and 4.5\% for $p$+A.

While most of the total entropy produced in a collision
typically comes from its initial stage,
a significant amount can potentially be produced during hydrodynamic evolution,
which in turn causes an increase in the charged hadron multiplicity.

\begin{figure}[htb]
\begin{center}
\includegraphics[width=9.0cm]{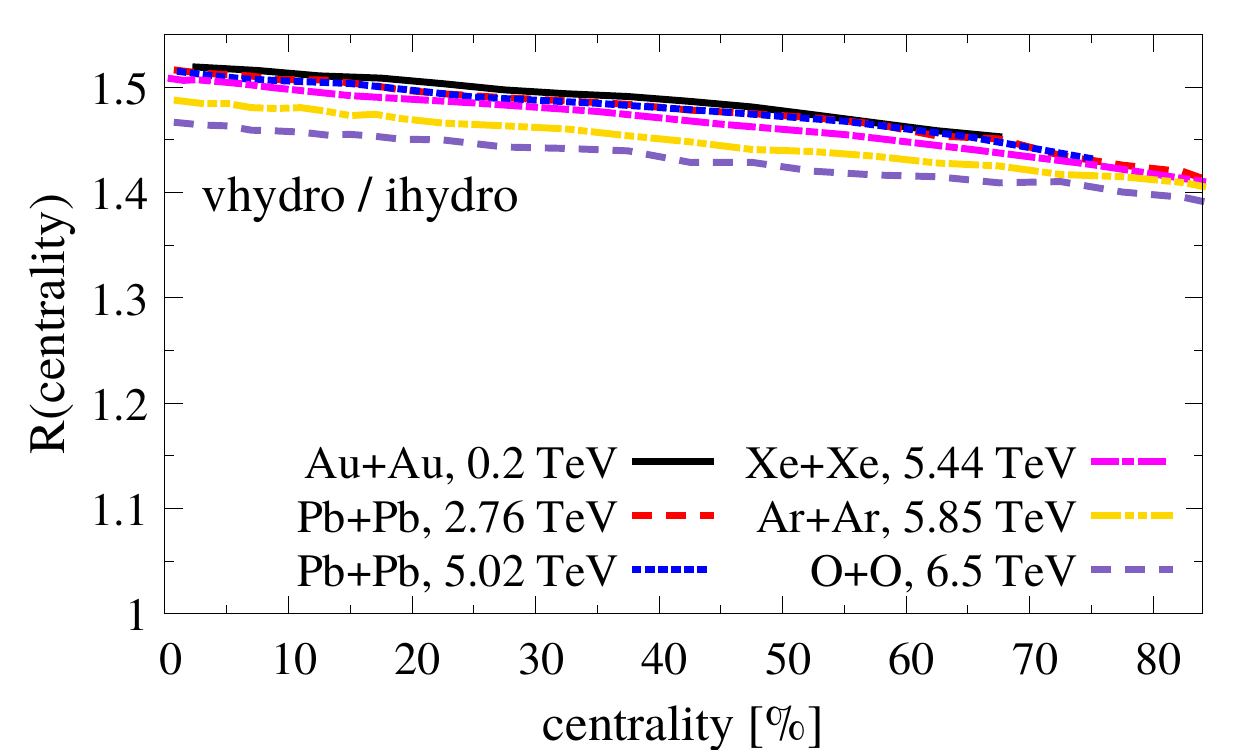}
\end{center}
\vspace*{-7mm}
\caption[a]{Centrality dependence of the ratio of our results with viscous
and ideal hydrodynamics for the charged hadron multiplicity and
its transverse energy distribution for selected collision systems.}
\label{fig:ratio_viscous_ideal}
\end{figure}

As entropy is exactly conserved in ideal hydrodynamics,
a ratio of the final hadron multiplicity generated
assuming ideal and viscous hydrodynamics evolution can be used
as a proxy to quantify the entropy which is produced on top of the
initial one after the particles that compose such system
stop interacting inelastically.
%%%
\fig{fig:ratio_viscous_ideal} shows the centrality dependence
of such ratio for several collision systems and energies.
As one can see, up to
$\sim$50\%
%$\sim$ 51\%--49\%
of the final multiplicity is produced from dissipative
effects happening during the late stages for the 0--10\% most central
heavy-ion collisions in the 0.2 TeV $\leq\sqrt{s}\leq$ 5.44 TeV region;
for intermediate collision systems such as Ar+Ar and O+O,
almost as much
%up to 48\%--45\% of
extra entropy is produced from hydro and final state effects in this same centrality
range. Although smaller systems might have a larger rate of entropy production,
in this case, their shorter lifetime (at similar energies) ensures that less total entropy is produced.
%%%
The production of entropy decreases quite slowly for
non-peripheral collisions and even collisions happening at 40\% of the
centrality range have between $\sim$ 49\%--42\%  of the final
multiplicity from hydro and final state effects (the first value is for Au+Au and
the second one is for O+O).
%%%

This hydrodynamic entropy production leads to an interesting observation: while we are able
to describe the data for global, bulk observables in both heavy-ion
and light-ion collision systems when hydrodynamic evolution is
either present or absent in both systems,
our results are incompatible with the case where it is included in A+A collisions,
but there is no hydrodynamic evolution in p+A collisions, as has been suggested
%\cite{Dusling:2012iga, Dusling:2017dqg, Dusling:2017aot, Mace:2018vwq, Mace:2018yvl, Mace:2019rtt}.
\cite{Dusling:2012iga, Dusling:2017aot, Mace:2018vwq}.
%[CITE].
In that scenario there would be a decrease in the overall
magnitude of charged hadron multiplicity in smaller systems
due to the absence of viscous entropy production.

\begin{figure}[htb]
\begin{center}
\includegraphics[width=8.5cm]{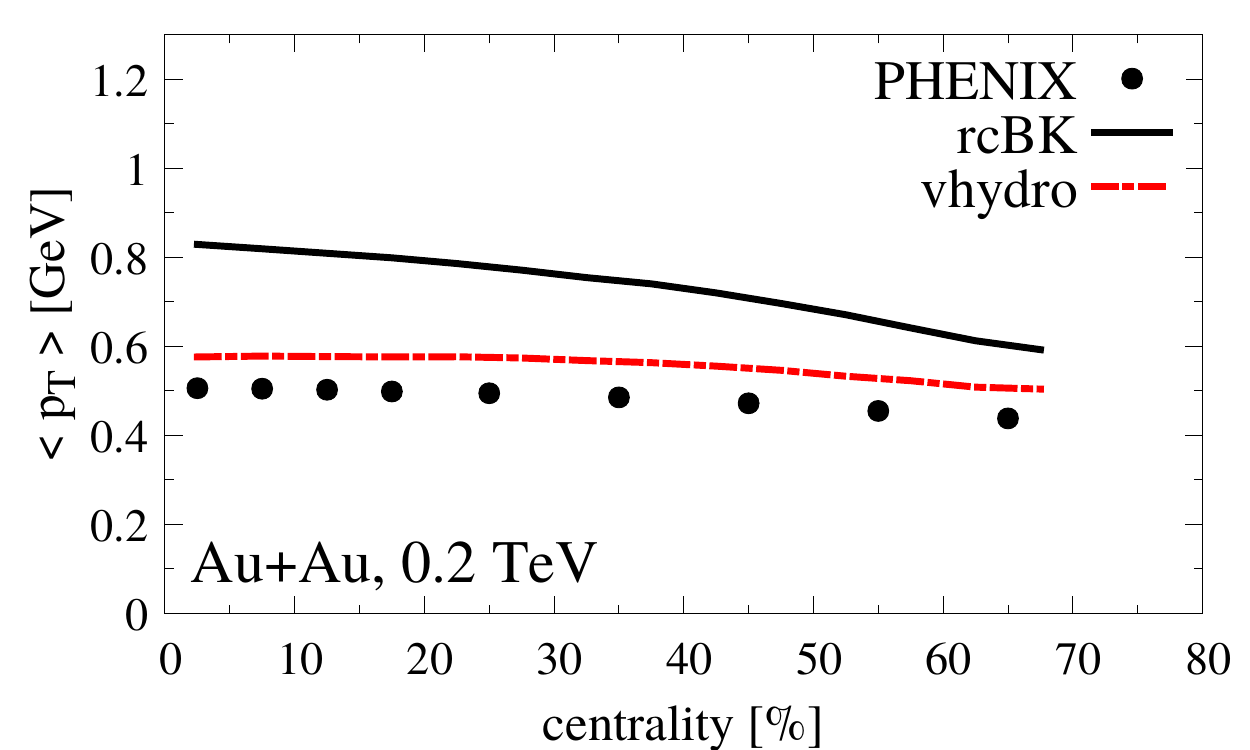}
\includegraphics[width=8.5cm]{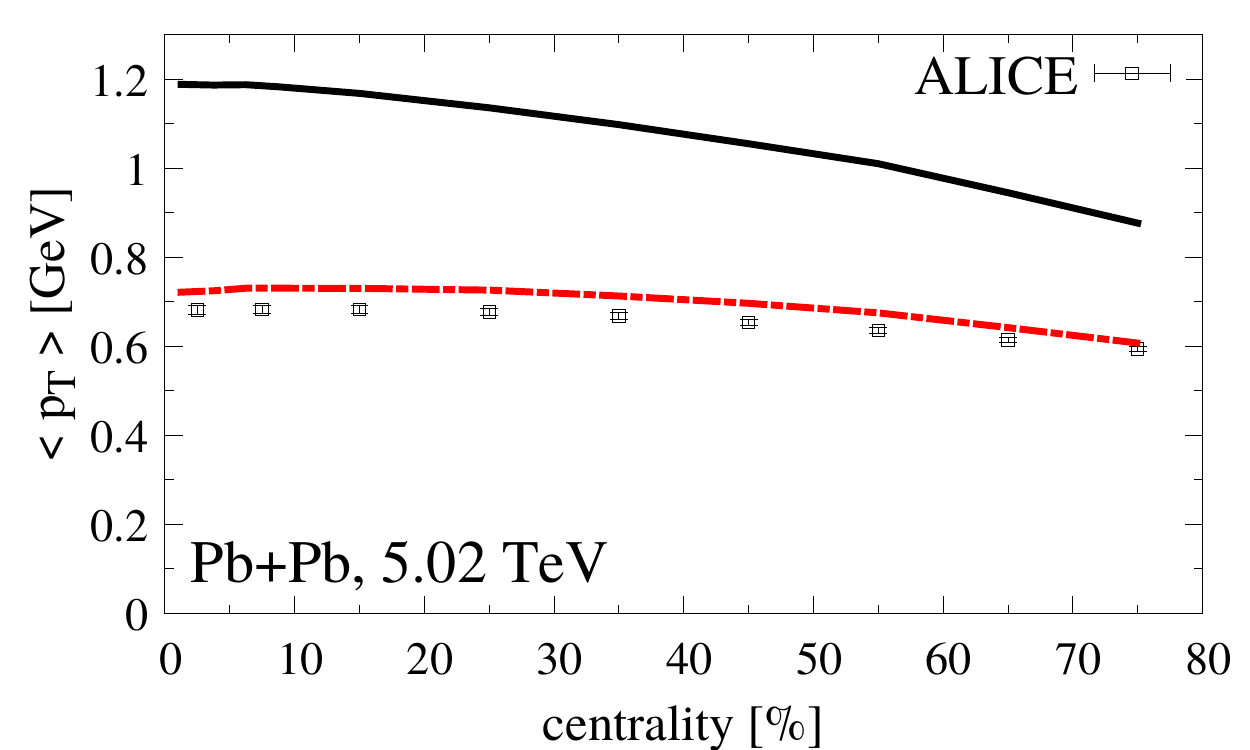}
\includegraphics[width=8.5cm]{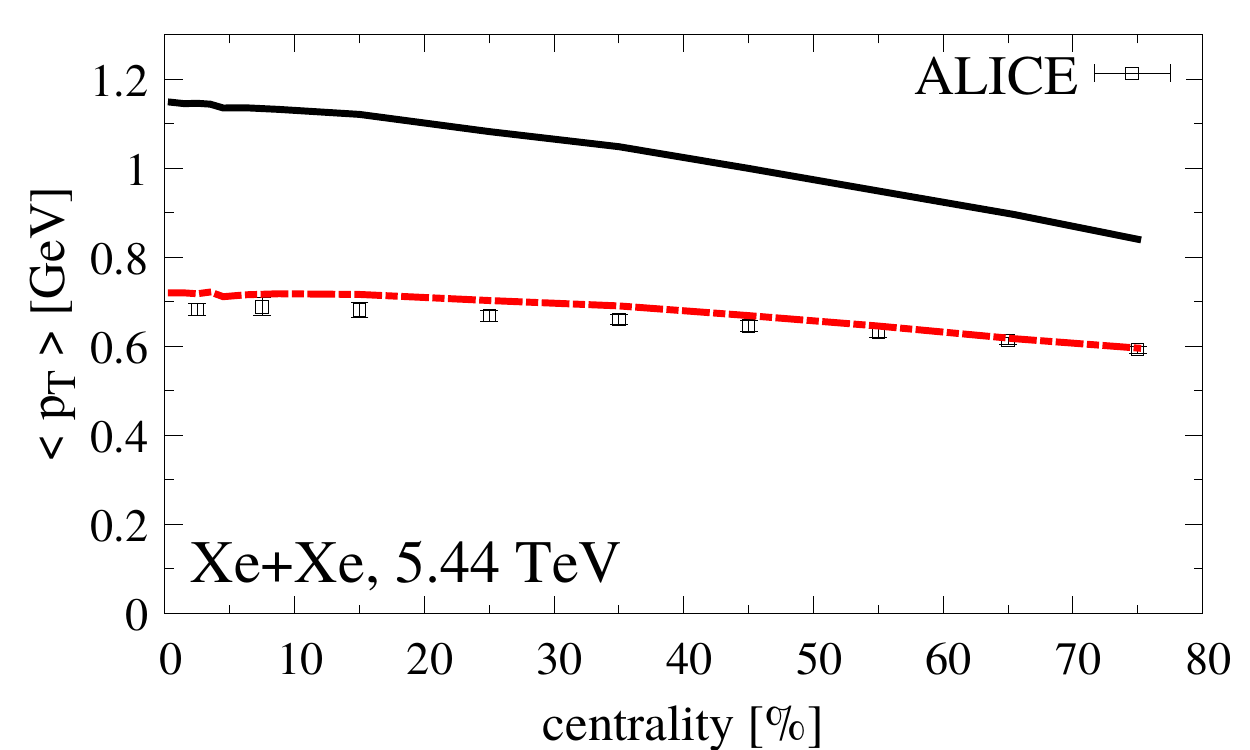}
\includegraphics[width=8.5cm]{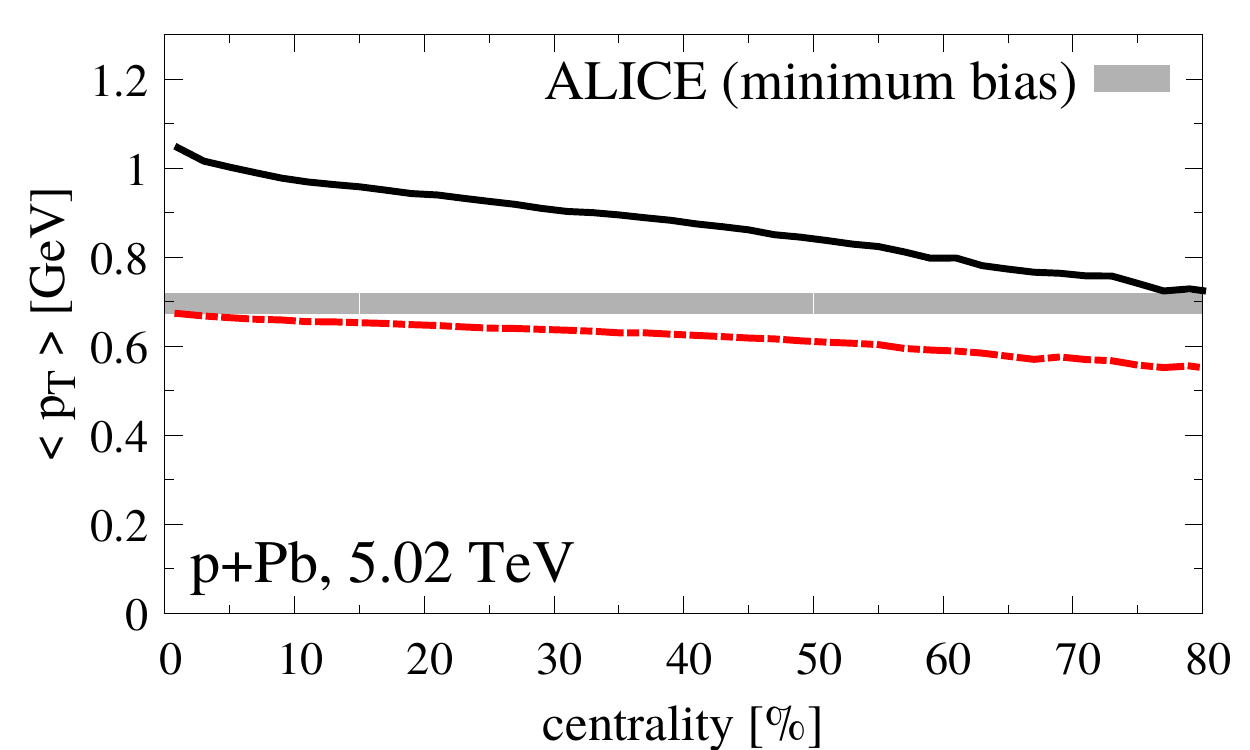}
\end{center}
\vspace*{-7mm}
\caption[a]{Centrality dependence of the average transverse momentum
of charged hadrons in Au+Au, Pb+Pb, Xe+Xe and p+Pb collisions.
The experimental data is from~\cite{Adler:2003cb,Abelev:2013bla,Acharya:2018eaq}.
}
\label{fig:avgpt}
\end{figure}
Next we consider the distribution of momentum, and how it is affected by hydrodynamic
evolution.
\fig{fig:avgpt} shows our results for the centrality dependence of the
average transverse momentum of charged hadrons at mid pseudorapidity
in heavy-ion collisions at RHIC\footnote{The PHENIX data~\cite{Adler:2003cb} at
200 GeV has originally been presented for identified particles (pions, protons and
kaons); in this case we loosely identify the sum of the average momentum of each
particle specie weighted by their relative fraction of the total particle
multiplicity as rough representation of the average transverse momentum of
(unidentified) charged hadrons.}
and LHC energies and p+Pb collisions also in the LHC regime.
%%%
The measurement has not been performed in $p$+Pb collisions
as a function of centrality or multiplicity, so we show the minimum
bias result~\cite{Abelev:2013bla} as a grey band in the bottom
right panel.

The rcBK results are significantly larger and show a faster increase
with the centrality of the collision with respect to the experimental data.
Such feature can be related to the increase of the saturation scale,
$Q_{s, {\rm nucl}}^{2}\sim N_{\rm part}\,Q_{s, {\rm proton}}^{2}$, together
with an effective ``free-streaming" space-time evolution of the system
leading to a final energy distribution per particle which is close
to the initial one.
%%%%

Unlike the case of multiplicity, whose centrality dependence is little changed
by hydrodynamic evolution,
the distribution of transverse
momentum is more sensitive to the dynamics happening during the evolution
of the system
%as it can be redistributed by medium and/or dissipative
%effects
\cite{Gyulassy:1983ub,Gyulassy:1997ib,Dumitru:2000up}.
%%%%
This fact is illustrated by the vhydro results, which show a
slower increase of $\langle p_T \rangle$ with centrality and are
much closer to the experimental data.

\begin{figure}[htb]
\begin{center}
\includegraphics[width=8.0cm]{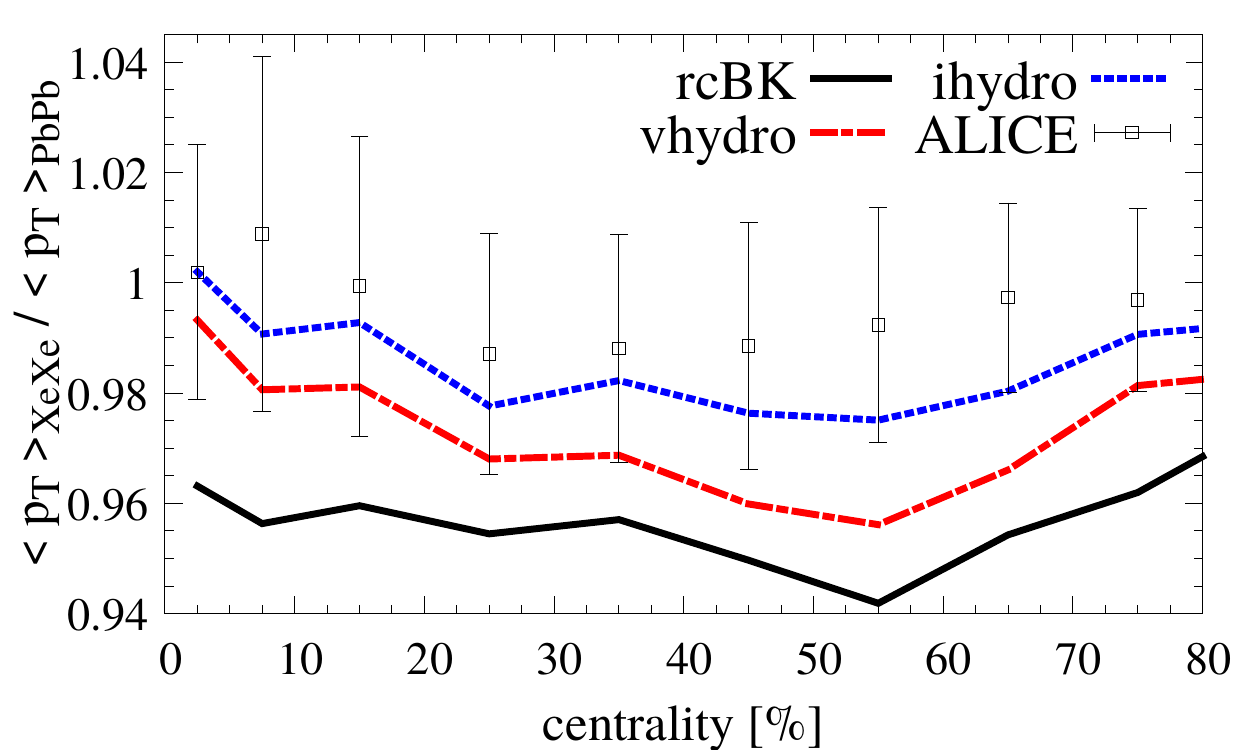}
\includegraphics[width=8.0cm]{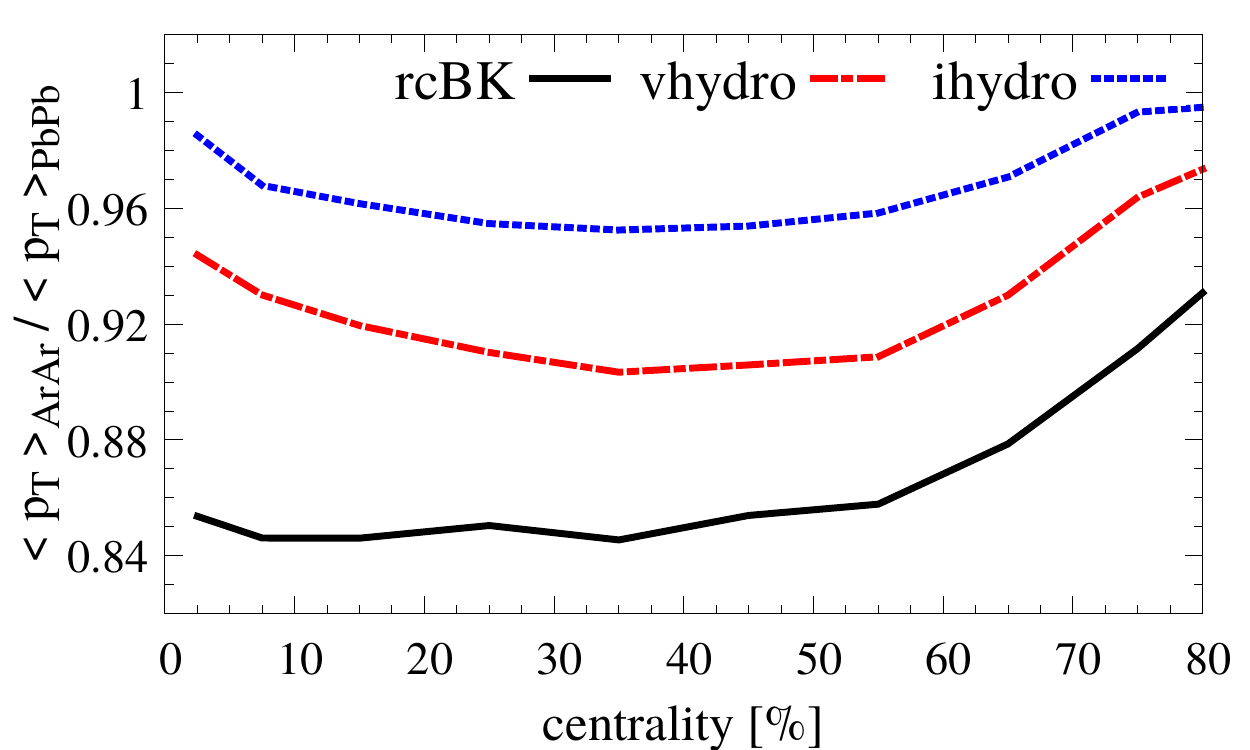}
\end{center}
\vspace*{-7mm}
\caption[a]{Centrality dependence of the ratio of the transverse energy
per charged particle from (left) Xe+Xe collisions at 5.44 TeV and Pb+Pb collisions at 5.02 TeV and
(right) Ar+Ar collisions at 5.85 TeV and Pb+Pb collisions at 5.02 TeV.}
\label{fig:avgpt_XeXeoverPbPb}
\end{figure}

It has been argued~\cite{Giacalone:2017dud} that the ratio of
mean transverse momentum in systems of different size
but at the same energy gives a robust test of hydrodynamic
behavior.   That is, it should depend little on the details
of a hydrodynamic system, but could be quite different
for a system with different dynamics.
To illustrate this, we show in \fig{fig:avgpt_XeXeoverPbPb}
(left) the ratio of
the average transverse momentum in Xe+Xe and Pb+Pb collisions
at 5.44 TeV and 5.02 TeV, respectively.

The scale invariance of ideal hydrodynamics
predicts that $\langle p_T \rangle$ should not
change with system size, broken only by
dissipative corrections.

Indeed, we see that our ideal hydrodynamic calculation gives
a ratio that is close to 1, in agreement with the recent measurement
from the ALICE Collaboration.   Our viscous calculation is somewhat
lower but is consistent
within error bars up to $\sim$40\% centrality.
Both results have the same shape
and differ by less than 2\%,
largely confirming the robustness
of the hydrodynamic prediction, though
perhaps indicating a potential probe of
viscosity.

On the other hand, the pure
initial state rcBK calculation is below the hydro prediction
and clearly incompatible with measured data.
%%%
We checked that this ratio decreases about 2\%, moving it
farther away from data (and the results of the hybrid
simulations), in case one considers the KLN UGD instead
of the rcBK one.
%%%
As  $\langle p_{T}\rangle \sim Q_{s,A}$ in the CGC framework and
$Q_{s,A}^{2}\sim A^{1/3}Q_{s,proton}^{2}$ for nuclear targets
(recall that $A^{1/3}$ is, roughly, the nuclear density probed by a projectile passing
through the center of a nucleus of mass number $A$),
the generic expectation would be
$\langle p_{T}\rangle_{XeXe}/\langle p_{T}\rangle_{PbPb}\sim (129/208)^{1/6}\sim 0.92$,
so that initial state models based on a $k_{T}$-factorized expression
would undershoot the data for the ratio of the
average transverse momentum in Xe+Xe and Pb+Pb collisions.
%%%
This measurement then imposes an important constraint on comparing
initial state models to observables involving the distribution of
energy between produced particles because even though much of the
uncertainty related to  the absolute value of the average transverse
momentum are canceled when taking a ratio, the pure CGC calculation
is still not in agreement with it in any centrality range.

We note that the split between each calculation (the pure initial state
and the hybrid simulation with ideal and viscous hydrodynamics)
becomes more apparent if one still keeps a similar collision energy but
increases the difference in system size with respect to Pb-Pb collisions.
%%%
This is demonstrated in the right panel of \fig{fig:avgpt_XeXeoverPbPb},
where now the average transverse momentum of charged hadrons
produced in Ar-Ar collisions is compared to the same quantity in
Pb-Pb collisions.
%%%%
This signals that the  onset of a hydrodynamic phase in heavy-ion collisions, along
with viscous effects, could, perhaps, be further investigated by studying the
centrality dependence of ratio of the mean $p_{T}$ across different collision
systems with similar collision energy.

\begin{figure}[htb]
\begin{center}
\includegraphics[width=8.0cm]{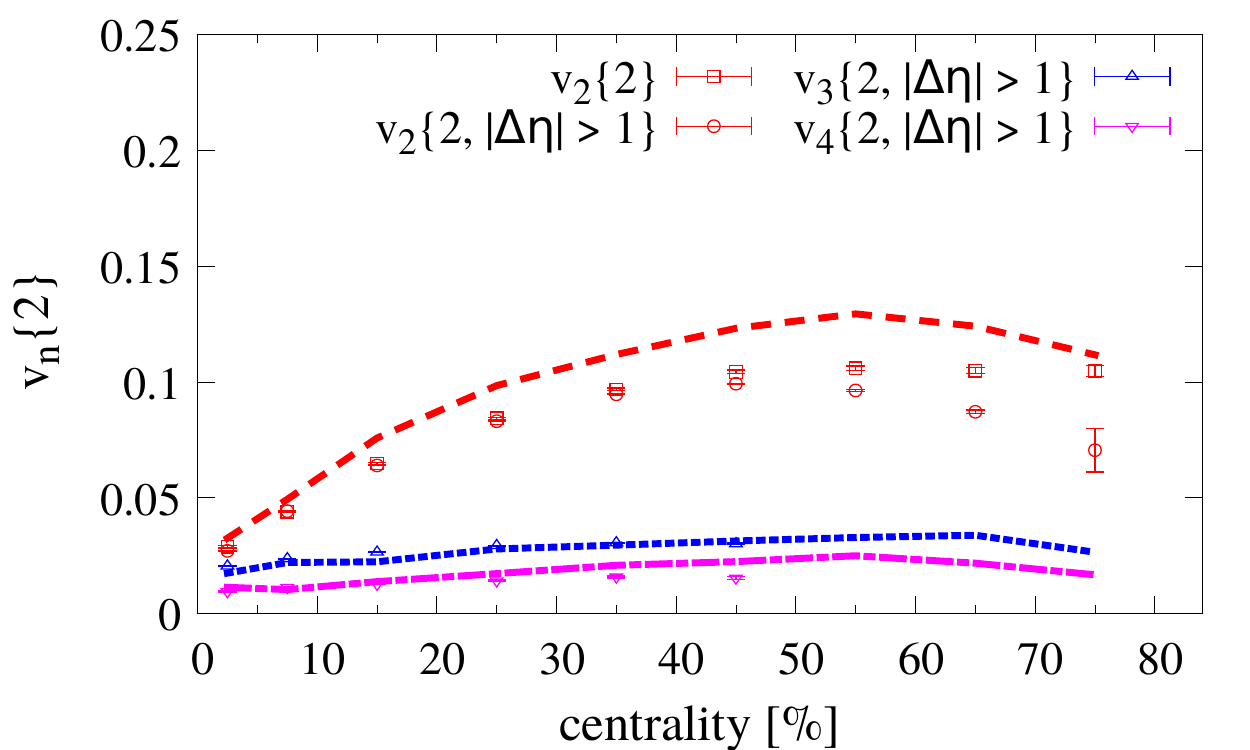}
\includegraphics[width=8.0cm]{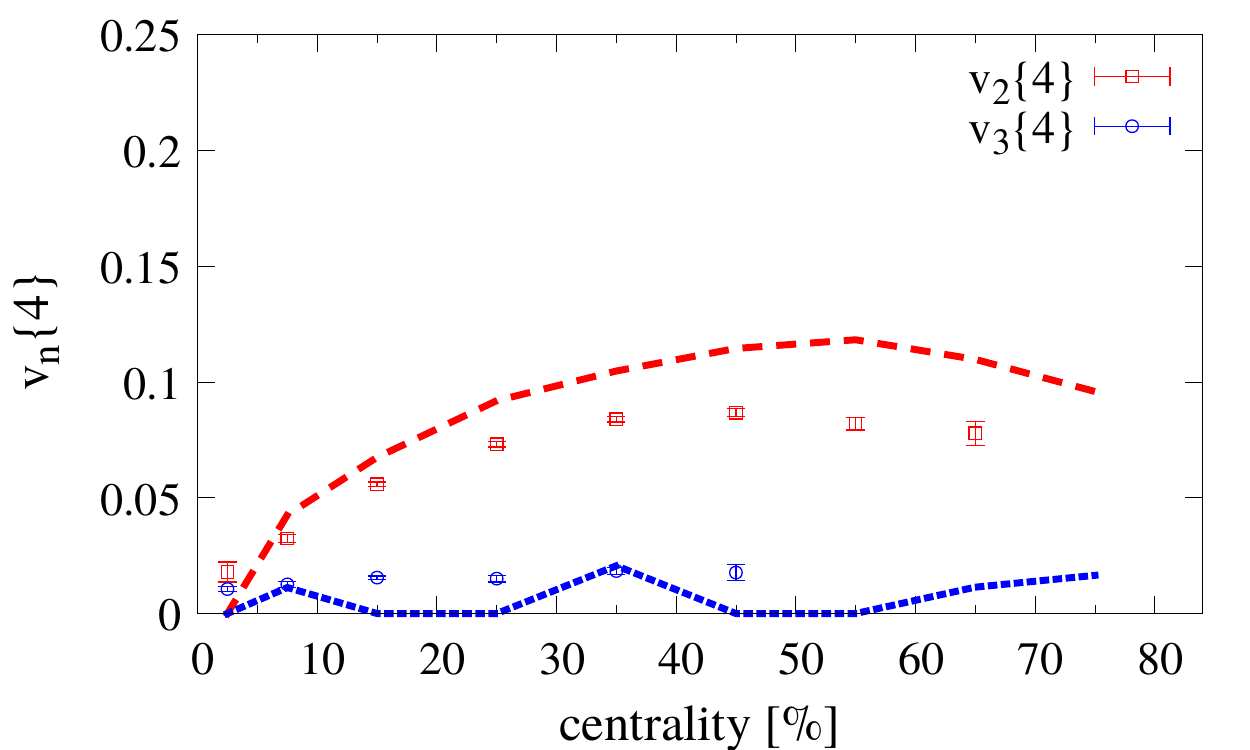}
\end{center}
\vspace*{-7mm}
\caption[a]{Centrality dependence of $v_{n}\{2\}$ and $v_{n}\{4\}$ in
Pb+Pb collisions at 2.76 TeV from the hybrid simulation. The experimental
data is from~\cite{Aamodt:2010pa,ALICE:2011ab}.}
\label{fig:vn2_and_vn4}
\end{figure}

Lastly, in \fig{fig:vn2_and_vn4} we compare the results
of our hybrid simulation to the centrality dependence of
the integrated n-th harmonic from 2- and 4-particle correlations,
$v_{n}\{2\}$ and $v_{n}\{4\}$ from Pb+Pb collisions at
LHC energies.
%%%
Despite the satisfactory agreement between the hybrid simulation
and the the bulk observables studied so far, we find that
the rcBK initial conditions from a running coupling improved
$k_{T}$-factorized expression still generate large
eccentricities (on average) which in turn lead to angular
asymmetries that are larger than the ones observed
experimentally by ALICE~\cite{Aamodt:2010pa,ALICE:2011ab}
collaboration.
As our hybrid model overshoots $v_{2}\{2\}$ while being in agreement
with $v_{3}\{2\}$ it
would be impossible to get a simultaneous
description of these harmonics for any value of viscosity with this
initial condition. This is in line with previous studies
which considered the leading order version of
\eq{eq:rcktfact}~\cite{Retinskaya:2013gca} where the running of the
coupling has been fixed by hand, as well as the previous
$k_{T}$-factorized MC-KLN model~\cite{Hirano:2009ah,Hirano:2010jg,Hirano:2010je}.
In all these cases, the average eccentricity of the early-time system is larger than
can be accommodated.

\section{Conclusions}

In this work we calculated
%global,
bulk observables measured
in the central rapidity region at different collider energies
from a CGC framework for particle production, and from a hybrid model
initialized by the same CGC calculation.
Due to the use of  local parton hadron duality
%the LPHD,
 in this simulation, the rcBK
dynamics represent all its dynamic content and determines
the shape of observables.
%%%
This approximation has been compared with a hybrid hydrodynamic
simulation, which accounts for medium and final state
effects, and which has been initialized using the same
CGC dynamics so that both approaches
have an intersecting point.
%%%

Assuming that fast thermalization occurs,
we estimate that up to $\sim$50\%
%49\%--42\%
of the final state
multiplicity observed in heavy-ion collisions can come from
dissipative effects during hydrodynamic evolution.

However, after fixing a single normalization factor in each of the two cases,
we find that that the
centrality dependence of the charged hadron
multiplicity is insensitive to the late
stage dynamics, matching the ones from the pure CGC
simulation for different collision systems and collision
energies in a wide centrality range. This fact is
also seen in the energy evolution of the charged
hadron multiplicity, where both simulations do not
differ more than 5\% in an energy range from
100 GeV to 10 TeV.

In contrast, the present framework does not accommodate
the case where hydrodynamic evolution is present in A+A
collisions but not p+A collisions.  In that case, the
entropy production in the large system can no longer be ignored
as an overall constant factor,
and the
system size dependence of multiplicity would be significantly stronger than seen
experimentally.

The evolution of the system and late stage dynamics
%however,
do play an important role in redistributing
momentum between the produced particles,
resulting in a
much better agreement with the measured
average transverse momentum.
%%%
We point out that comparing the average transverse
momentum in Pb-Pb collisions to the same quantity
in other colliding systems (at similar energies)
as a function of centrality could be
considered for two purposes: to investigate the onset
of hydrodynamical phase in high-energy heavy-ion
collisions
%%%  since the difference of initial state and hybrid simulations increase with the difference of system size
and to probe the effects of viscosity in different
colliding systems.

Finally, we verified that the running coupling corrections
encoded in \eq{eq:rcktfact} do not change previous results
for harmonic flow coefficients~\cite{Retinskaya:2013gca}
where a leading order expression with the running of the
coupling fixed by hand was used, and a large relative value
 $\varepsilon_2/\varepsilon_3$ prevents good agreement
with measured data of elliptic and triangular flow.

\begin{acknowledgments}
We thank Jorge Noronha for useful discussions and Marco van Leeuwen
for informations related with ALICE data.
A.V.G. thanks Yu. Kovchegov for useful email exchange and
acknowledges the Brazilian funding agency FAPESP
for financial support through grant 2017/14974-8.
%%%
F.G.~acknowledges  support  from Conselho Nacional de Desenvolvimento
Cient\'{\i}fico e Tecnol\'ogico (CNPq grant 310141/2016-8).
%%%
M.L.~acknowledges support from FAPESP projects 2016/24029-6 and 2017/05685-2.
%%%
F.G. and  M.L.  acknowledge support from project INCT-FNA Proc.~No.~464898/2014-5.
%%%
A.V.G. acknowledges the use of resources of High Performance
Computing made available by Superintend\^encia de Tecnologia
da Informa\c{c}\~ao da Universidade de S\~ao Paulo.
\end{acknowledgments}

\end{document}